\documentclass[referee]{raa}            
\usepackage{graphicx,times}             
\usepackage{natbib}
\usepackage{amssymb,amsmath}
\bibpunct{(}{)}{;}{a}{}{,}
\usepackage[a4paper=true,dvipdfm=true,pagebackref=true]{hyperref}
\hypersetup{colorlinks = true, linkcolor = green, anchorcolor = red, citecolor = blue, filecolor = red, pagecolor = red, urlcolor = red}

\begin{document}

   \title{On the Formation of the Double Neutron Star Binary PSR J1846-0513
\,$^*$
\footnotetext{$*$ Supported by the National Natural Science Foundation of China.}
}

   \volnopage{Vol.0 (20xx) No.0, 000--000}      
   \setcounter{page}{1}          

   \author{Long Jiang
      \inst{1,4,6}
   \and Kun Xu
      \inst{1}
   \and Shuai Zha
      \inst{2}
   \and Yun-Lang Guo
      \inst{3}
   \and Jian-Ping Yuan
      \inst{4}
   \and Xiang-Li Qian
      \inst{5}
   \and Wen-Cong Chen
      \inst{1,6}
   \and Na Wang
      \inst{4}
   }

   \institute{School of Science, Qingdao University of Technology,
             Qingdao 266525, China; {\it chenwc@pku.edu.cn}\\
        \and
             Yunnan Observatories, Chinese Academy of Sciences,
             Kunming 650216, China; \\
        \and
             Department of Astronomy, Nanjing University,
             Nanjing 210023, China; \\
        \and
             Xinjiang Astronomical Observatory,
             Chinese Academy of Sciences, Urumqi 830011, China; {\it na.wang@xao.ac.cn}\\
        \and
             Department of Intelligent Engineering,
             Shandong Management University, Jinan 250357, China;\\
        \and
             School of Physics and Electrical Information,
             Shangqiu Normal University, Shangqiu 476000, China\\
\vs\no
   {\small Received~~20xx month day; accepted~~20xx~~month day}}

\abstract{ The double neutron star PSR J1846-0513 is discovered by the Five-hundred-meter Aperture
Spherical radio Telescope (FAST) in Commensal Radio Astronomy FAST Survey.
The pulsar is revealed to be harbored in an eccentric orbit
with $e=0.208$ and orbital period of 0.613 days.
The total mass of the system is constrained to be $2.6287(35)\rm{M}_{\odot}$,
with a mass upper limit of $1.3455{\rm~M}_{\odot}$ for the pulsar and a mass lower limit of $1.2845{\rm~M}_{\odot}$ for the companion star.
To reproduce its evolution history, we perform a 1D model for the formation of PSR J1846-0513 whose progenitor is assumed to be neutron star - helium (He) star system via MESA code. Since the large eccentricity is widely believed to originate from an asymmetric supernova explosion,
we also investigate the dynamical effects of the supernova explosion. Our simulated results show that the progenitor of PSR J1846-0513 could be a binary system consisting of a He star of $3.3-4.0{\rm~M}_\odot$ and a neutron star in a circular orbit with an initial period of $\sim0.5$ days.
}

   \authorrunning{L. Jiang, K. Xu, et al. }            
   \titlerunning{Formation of DNS PSR J1846-0513}  

   \maketitle

%
%
\section{Introduction}           
\label{sect:intro}

As evolutionary productions of massive binary stars, double neutron star (DNS) systems not only are
ideal probes in understanding stellar and binary evolution and supernova (SN) explosions (\citealt{van76}),
but also provide many chances testing astrophysicsal and fundamental physical laws.
Firstly, the first-born neutron star (NS) in a DNS is a (mildly) recycled pulsar,
which originates from an accretion of material and angular momentum form the donor star (\citealt{tau06} and references therein).
Secondly, the orbital evolution of close DNS are widely used to test gravitational theories (\citealt{wei10,fon14} and references therein). Thirdly, the merger of DNS systems is generally thought to be responsible for the origin of some short gamma-ray bursts, kilo-nova, high-frequency gravitational wave (GW) events (\citealt{li98,xia17,yu18}).
Fourthly, the GW events producing by DNS mergers are the most important targets of multi-messenger astrophysics, and provide a channel understanding the origin of the matter with over nuclear density (\citealt{yu19} and references therein). Finally, DNS systems are ideal objects to accurately measure the masses of NSs, which can be used to constrain the equation of
state with high-density nuclear matter (\citealt{ozel16}).

The Hulse-Taylor pulsar (PSR B1913+16) is a famous DNS system,
which includes a radio pulsar with a spin period of $59\rm~ms$ (\citealt{hul75}).
The two components orbit each other in an eccentric orbit with an eccentricity of 0.617 and an orbital period of 0.323 days.
The detailed timing analysis for PSR B1913+16 indirectly confirmed the prediction of GW radiation (\citealt{wei10} and references therein).
Assuming a 'kick' exerted on the newborn NS due to an asymmetric SN explosion,
\cite{fla75} found that a 'kick' velocity of $200{\rm~km~s}^{-1}$ can lead to the detected eccentricity.
Considering that the radio pulsar is an old recycled pulsar, \cite{van76} gives the first evolutionary scheme from zero-age-main-sequence (ZAMS) massive binary to the formation of DNS. After about half a century, it is widely accepted as standard scheme of isolated massive binary evolution and DNS formation.
\cite{hil83} conducts a comprehensive analytical study on the dynamic effects exerted on binary orbits by asymmetric SN.
These formula are widely used to study the SN with given 'kick' velocity, especially with Monte Carlo simulations and statistical analysis.
Utilizing a Maxwellian distribution of kick velocities,
\cite{kal96} derives analytical expressions for the distribution functions of post-SN orbital parameters,
including eccentricity and separation which is an important development of the work of \cite{hil83}.
Assuming that the asymmetric SN with a high kick velocity may disrupt the binary,
\cite{tau98} deduces the runaway velocities of the two components.  Their results give clues to the origin of the proper motion of some isolated (mildly) recycled pulsars.

Using the Monte Carlo simulations, \cite{bra95} systematically studied the influence of SN kick
velocities on the orbital parameters of post-SN NS binaries, and obtained the post-SN distributions of orbital parameters for progenitors of high-mass X-ray binaries and low-mass X-ray binaries. They found that the 'kick' velocity of SN plays an important role to the evolution and results of both high-mass X-ray binaries and low-mass X-ray binaries. Utilizing the self-adaptive, non-Lagrangian mesh-spacing Eggleton stellar-evolution code (\citealt{egg71,egg72, egg73,pol95}), \cite{dew02} calculated the evolution of NS binaries with a ZAMS helium star in the mass range of $(1.5-6.7){\rm~M}_\odot$ and a NS of $1.4~{\rm M}_\odot$,
and found that the Case BB mass transfer from helium stars more than $2.6-2.8~{\rm M}_\odot$ will produce DNS systems.

Employing the BEC stellar evolution code (\citealt{lan98, wel01, yoo10}),
\cite{tau13} modeled the evolution of a binary system consisting of a helium star of $2.9{\rm~M}_\odot$ and a NS of $1.35{\rm~M}_\odot$ in an orbit with an initial period of 0.1 day until the He star evolves to off-center oxygen burning ignition. Comparing with the results of \cite{ume12} and \cite{jon13}, \cite{tau13} found the chemical abundance structure of their simulated core is similar to that of the core of single stars with a masses of $(9.5-11)\rm{M}_{\odot}$ prior to core collapse SN (CCSN).
Subsequently, \cite{tau15} performed a systematic investigation on the evolution of post-common envelope (CE) binaries including a naked helium star with masses of $(2.5-10.0)~\rm{M}_{\odot}$ and a NS with a mass of $1.35~\rm{M}_{\odot}$ in an orbit with initial orbital periods of $0.6-120$ days, and a detailed study on the expected properties of the ultra-stripped SN explosion.
Based on these results and the observed parameters of known DNS systems, \cite{tau17} found empirical evidence to support the hypothesis that ultra-stripped SNe with small kicks and larger kicks generally produce less massive NSs of $\la 1.3{\rm~M}_\odot$ and more massive NSs of $> 1.3{\rm~M}_\odot$, respectively. Mimicking binary evolution in single-star evolution by treating mass loss, \cite{mor17} successfully modeled the evolution of zero-age main-sequence (ZAMS) He star until the iron core infall at a velocity of $1000{\rm~km~s}^{-1}$.
Synthesized light curves and spectra of ultra-stripped supernovae, they suggested typical peak luminosities around $10^{42}{\rm~erg~s}^{-1}$. Comparing synthesized and observed spectra, they found that SN 2005ek, and SN 2010X may be related to ultra-stripped supernovae.

Because the mass exchange and the tidal forces between two components may change the stellar nuclear burning scheme,
the final core masses of He stars in close binaries should be significantly different from those in single star evolution (\citealt{tau13}).
\cite{jia21} conducted the first detailed and self-consistent 1D stellar evolution model of a post-CE binary consisting of a $3.20~M_\odot$ ZAMS He star and a $1.35~M_\odot$ NS in an orbit with an initial orbital period of
1.0 day until ultra-stripped Fe CCSN, i.e., the infall velocity of the Fe core exceeds $1000{\rm~km~s}^{-1}$.
Their simulated post-SN system can merge with a probability of 30\% within a Hubble time, potentially generating a GW event like GW170817.

Once the degenerate oxygen-neon (ONe) core mass increases to the Chandrasekhar limit,
the electron-capture (EC) process on $^{24}\rm Mg$ and $^{20}\rm Ne$ may trigger a EC-SN event, and result in the formation a NS (\citealt{nom84,nom87,sie07,jon13,tau15,tau23}). Considering that the explosive oxygen burning may occur in the near-Chandrasekhar mass ONe core,
\cite{guo24} found that NS+He star systems can evolve toward most of the observed DNS systems,
and obtained initial parameter spaces in the initial He star mass versus initial orbital period plane.
Their results show that, with kick velocity of $50{\rm~km~s}^{-1}$,
the majority of the observed DNS systems can be explained with the EC-SN channel, and half of the
observed DNS systems seem to originate from tight pre-SN orbit with orbital period less than 1 day.

Recently, the Commensal Radio Astronomy FAST survey pulsar search observations
has confirmed the discovery of $\sim200$ new pulsars including 40 MSPs and 3 DNS candidates including
J2150+3427 (\citealt{wu23}), J1901+0658 (\citealt{su24}), and J1846-0513(J1846, here after) (\citealt{zhao24}).
The spin period and spin-down rate of J1846 are $23.36~\rm ms$ and $1.0106(3)\times10^{18}~\rm s\,s^{-1}$, respectively. The characteristic age and surface magnetic field are derived to be $0.3662~\rm{Gyr}$ and $4.92\times10^{9}\rm{G}$ respectively,
appearing as the properties of a recycled pulsar. The source locates in an orbit with an orbital period of $P_{\rm{orb}}=0.613~\rm{day}$ and an eccentricity of $e=0.208$. The measurement of the periastron advance constrained the total mass to be $2.6287(35)~\rm{M}_{\odot}$, with an upper mass limit of $1.3455{\rm~M}_{\odot}$ for the pulsar
and a lower mass limit of $1.2845{\rm~M}_{\odot}$ for the companion. Some binary parameters including the mass ratio, the distributions of spin period and orbital period, and the relationship between orbital period and eccentricity are consistent with those of the recycled pulsars in known DNS systems (\citealt{zhao24}).

Based on the extensive distributions of known DNS in the eccentricity-orbital period diagram, the eccentricity-second NS mass diagram, and the NS star mass distribution (\citealt{tau17}), it is prudent to investigate the DNS systems with several different isolated binary evolution models. In this work, we perform a comprehensive stellar evolution model until an ultra-stripped iron core collapse for the formation of J1846. Considering its similarities with B$1546+16$ (\citealt{fon14}), and J$1756-2251$ (\citealt{fau05, fer14}), specifically, second NS mass of $\sim1.2-1.3~\rm{M}_{\odot}$, orbital eccentricity of $\sim0.2-0.3$, and orbital period of $\sim0.3-0.6~\rm{day}$, our results may provide valuable insights into their evolutions.

\section{Methods}
\label{sect:Obs}

According to the standard formation scenario of DNS (\citealt{van76, bha91, tau17, tau23}),
a ZAMS binary system with two O/B stars experiences the first stable mass transfer from the more massive primary to the less massive secondary via Roche-Lobe overflow (RLO). After the hydrogen-rich envelope is stripped, the primary star evolves into a helium star,
then forms a NS via a Type Ib/c SN (\citealt{yoo10}). Subsequently, the massive secondary star fills its Roche Lobe and transfers the material onto the NS. Consequently, the system appears as a high-mass X-ray binary (HMXB) (\citealt{liu00}).
Due to an extreme mass ratio, the mass transfer becomes unstable after the secondary star climbs to the giant branch,
and the runaway mass transfer results in the system to enter a CE evolution phase (\citealt{pac76,iva13}).
If the CE can be successfully ejected, a compact system consisting of a NS and a naked He star is produced, then a stable Case BB mass transfer initiates due to the expansion of the He star (\citealt{hab86,dew02,tau15}).
By accreting the material and angular momentum form the He star, the NS is spun up to a spin period of $\sim 10-100$ ms \footnote {The acceleration during previous HMXB and CE is still under debate, see \cite{tau17}, and references therein},
while the He star experiences
an ultra-stripped SN explosion and evolves into a new-born NS (\citealt{tau13,tau15,suw15,mor17,mul18,mul19})
or EC-SN (\citealt{tau17, guo24} and references therein).

To reproduce the evolutionary history of J1846, we simulate its evolution via the $\sc MESA$ code module $\sc MESAbinary$ of version r-12778~(\citealt{pax11,pax13,pax15,pax18,pax19} and references therein).
The beginning of our simulation is post-CE binary systems consisting of a NS and a He star in a circular orbit.
To fit the observed mass upper limit ($1.3455{\rm~M}_{\odot}$) of the pulsar , the NS is considered as a point mass with an initial mass of $1.345{\rm~M}_{\odot}$. The companion star is assumed to be a naked He ZAMS star with a chemical composition of  $Y = 0.98$ and $Z = 0.02$.
The initial mass $M_{\rm~He,i}$ and orbital period $P_{\rm~orb,i}$ of three systems are listed in Table 1.
Nuclear network $\sc mesa235.net$ is adopted
which includes 235 nuclei from $^1{\rm H}$ to $^{73}{\rm Ge}$.
Following \cite{lan91}, we take the mixing length parameter as $\alpha=l/H_{\rm{p}}=1.5$,
where $l$ and $H_{\rm{p}}$ are the mixing length and local pressure scale, respectively.
We calculate the mass loss rate by the stellar winds using the 'Dutch' scheme with a scaling factor of 1.0 in $\sc MESA$, which comes from some works performed by several Dutch authors (\citealt{gle09,vin01,nug00}). The Type 2 opacities ( from the OPAL tables, \citealt{igl93}) are used to calculate the extra
C/O burning during and after He burning.
Ignoring the wind accretion onto the NS, we assume that the wind leaves the system from the surface of the He star,
carrying away its specific orbital angular momentum (AM).

Due to the narrow orbital separation, the He star will fill its Roche lobe,
resulting in the transfer of He-rich material to the NS via the inner Lagrangian point.
To calculate the mass transfer rate, we employ the 'Kolb' scheme (\citealt{kol90}) based on optical thick overflow.
Since the mass-transfer rate significantly exceeds the Eddington accretion rate limit of the NS ($|\dot{M}_{\rm He}|>>\dot{M}_{\rm Edd}$), most of the transferred matter was ejected from the vicinity of the NS as isotropic fast wind, taking away its specific orbital AM (\citealt{bha91, tau06, tau23}). The Eddington accretion rate was fixed to be $\dot{M}_{\rm Edd}=4.0\times10^{-8}~{\rm M_\odot~yr}^{-1}$ for helium material since the mass increase of the NS is ignorable, $<0.01~{\rm M_\odot}$, during the short mass transfer timescale of RLO.
Furthermore, when the evolution time-step exceeds 1.0 yr,
the element diffusion caused by convention is also considered, based on the results of \cite{sta16}.

During the evolution from He ZAMS to Fe CCSN, the temperature, density, and pressure in the central region will vary by several magnitudes.
Therefore we divided the simulation into three stages, consistent with the study of \citealt{jia23}):
(i) Form He ZAMS to the end of carbon burning, prior to the ignition of neon.
In this stage, the central temperature climbs up to $T_c\sim10^9{\rm K}$,
while the mass fraction of carbon drops to $0.2\%$.
(ii) Evolution continues to $T_c\sim10^{9.85}{\rm K}$,
where the core masses of silicon and iron raise up to their maximum,
and the Fe core infall starts, minutes prior to CCSN.
(iii) Evolution progresses to Fe CCSN,
in which the Fe core infall velocity approaches $1000{\rm~km~s}^{-1}$ or
the time-step declines to $1.0\times10^{-7}\rm~s$.

\begin{table}[h]
\begin{center}
\caption[]{ Initial and Final binary Parameters of Simulated Systems}\label{Tab:publ-works}
 \begin{tabular}{cllllll}
  \hline\noalign{\smallskip}
Sys & $M_{\rm NS,i}$  & $M_{\rm He,i}$   & $P_{\rm orb,i}$  & $M_{\rm NS,f}$   & $M_{\rm He,f}$  & $P_{\rm orb,f}$ \\
    & ${\rm M}_\odot$ & ${\rm M}_\odot$  &  day             & ${\rm M}_\odot$  & ${\rm M}_\odot$ &  day    \\
  \hline\noalign{\smallskip}
A  & 1.345 & 3.3  & 0.55  & 1.3455 &  1.90 & 0.410	          \\ 
B  & 1.345 & 3.5  & 0.50  & 1.3454 &  2.08 & 0.418	          \\
C  & 1.345 & 4.0  & 0.50  & 1.3452 &  2.63 & 0.397 	       \\
  \noalign{\smallskip}\hline
\end{tabular}
\end{center}
\end{table}

\section{Results}
The initial and final parameters of three simulated systems are listed in Table 1,
while some key epochs in their evolutions are illustrated in Table 2.
$t_{\ast}~(\sim1-2{\rm~Myr})$\footnote {The initial age of the He ZAMS is set to zero.}
is the age of He star at CCSN, i.e.,
the lifetime of the He star, and $(t_{\ast}-t)$ is the remaining time.
In summary, in our systems, the He stars expend $\sim90\%$
of their lifetime before RLO, and lost $\sim1.4{\rm~M}_\odot$ material in $\sim10^4{\rm ~yr}$.
During their evolution, the NS accrete $*10^{-4}{\rm~M}_\odot$,
and the binary orbital periods evolve to about 0.4 day.
Furthermore, the duration of Stage 2 is about $(1.3-2.5)\times10^3{\rm ~yr}$ for Sys. A and B,
while it is only $\sim20{\rm ~yr}$ for Sys. C.
Finally, the timescales of three systems in the 3rd stages are similar ($3-4{\rm~min}$).

\begin{table}[h]
\begin{center}
\caption[]{Key Epochs in the Simulations}\label{Tab:publ-works}
 \begin{tabular}{cccccccc}
  \hline\noalign{\smallskip}
Sys & Onset of Case BB RLO    & Start Point of Stage 2  &  Start Point of Stage 2  &  Fe-CCSN  \\
    & $t_{1}$                 & $t_{2}$                 & $t_{3}$                  & $t_{\ast}$\\
    & $t_{\ast}-t_{1}$        & $t_{\ast}-t_{2}$        & $t_{\ast}-t_{3}$         &  \\
  \hline\noalign{\smallskip}
A   & $1.69{\rm~Myr}$         & $1.70249{\rm~Myr}$      &  $\sim t_{\ast}$         & $1.702496352223{\rm~Myr}$	     \\
    & $\sim16000{\rm~yr}$     & $\sim2500{\rm~yr}$      &  $\sim3.0{\rm~min}$                       \\
B   & $1.56{\rm~Myr}$         & $1.57278{\rm~Myr}$      &  $\sim t_{\ast}$         & $1.574060120323{\rm~Myr}$  	\\
    & $\sim13000{\rm~yr}$     & $\sim1300{\rm~yr}$      &  $\sim2.7{\rm~min}$                       \\
C   & $1.32{\rm~Myr}$         & $1.32743{\rm~Myr}$      &  $\sim t_{\ast}$         & $1.327456558025{\rm~Myr}$ \\
    & $\sim6000{\rm~yr}$      & $ \sim~20{\rm~yr}$      &  $\sim3.8{\rm~min}$                        \\
  \noalign{\smallskip}\hline
\end{tabular}
\end{center}
\end{table}

Fig. 1 plots the evolution of orbital periods for our simulated three systems.
It is clear that the orbital evolution of these three systems is very similar.
Before the onset of Case BB RLO, the orbital periods of three systems gradually increase to maximum values due to a strong wind loss from the He stars. Subsequently, the orbital periods significantly decrease to a value of 0.4-0.42 days.
Considering the evolutionary similarity of three systems, we only present the simulated results of Sys. A in the following figures.

 \begin{figure}[h]
   \centering
   \includegraphics[trim=10 250 100 30, width=\textwidth, angle=0]{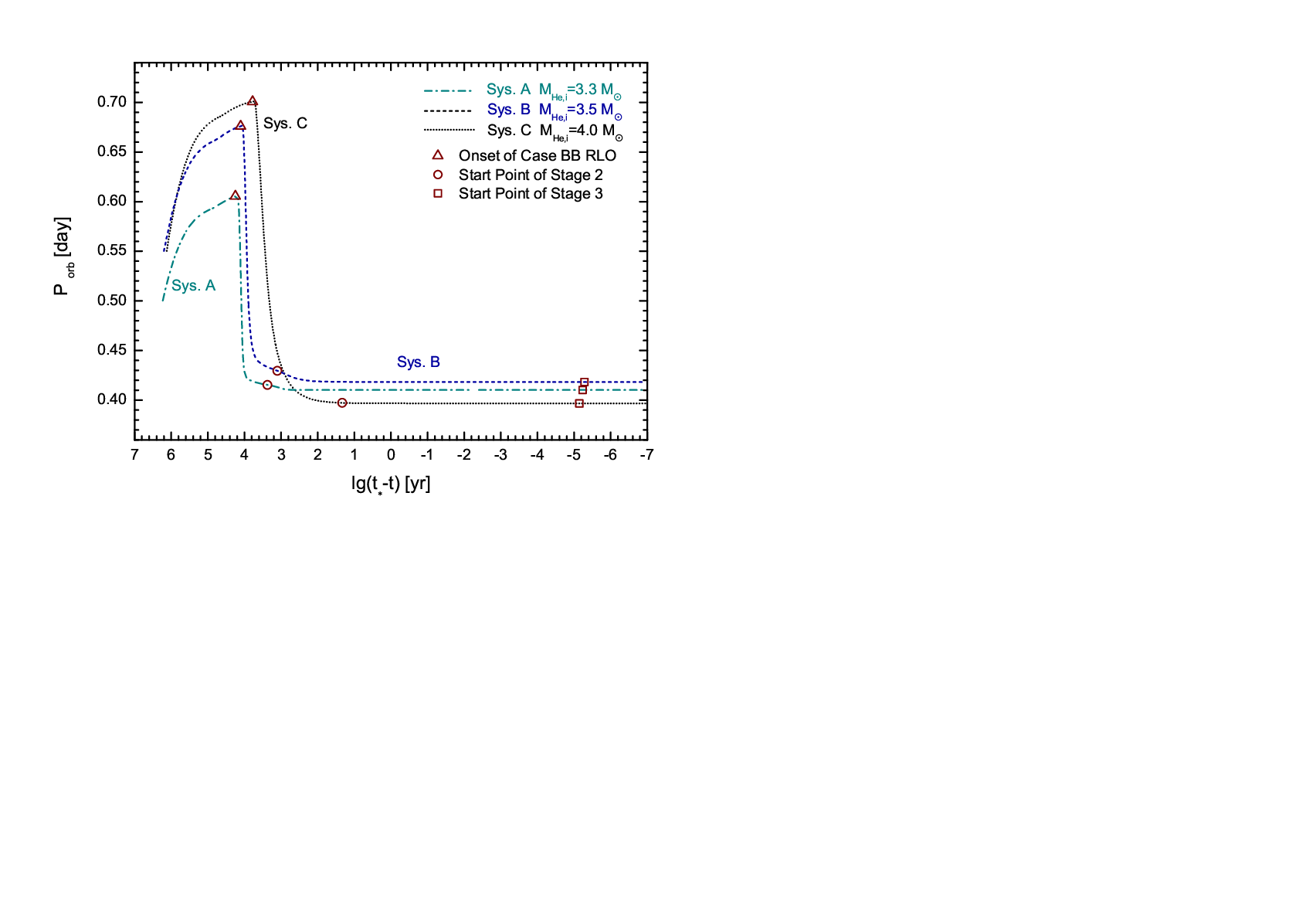}
   \caption{Evolution of orbital periods vs. time remaining until CCSN. The evolutionary tracks of systems A, B, and C are represented by the dot-dashed dark cyan, short dashed blue, and short dotted black curves, respectively.
   The triangles, circles, and squares correspond to the onset of Case BB RLO, the start points of stage 2, and 3, respectively. }
   \label{Fig1}
   \end{figure}

To comprehend the orbital evolution, we show the evolution of the mass loss rates of the He stars in Fig. 2.
The onset of RLO and the starting points of stage 2 and 3 are marked with vertical gray dot-dash, dash, and dot lines, respectively. The horizontal solid gray line represents the Eddington accretion rate ($\dot{M}_{\rm Edd}=4.0\times10^{-8}~M_\odot\,\rm yr^{-1}$) of the NS.
The vertical gray short dashed lines marked with '$M_{\rm Si}>0$' at ${\rm lg}(t_\ast-t)\sim1.2$ indicates the formation of a silicon (Si) core. Its relationship to the fluctuations of mass loss rates driven by RLO and wind will be discussed comprehensively at the end of this section. The mass loss rates due to the stellar winds and RLO are illustrated by the dot-dashed purple and dotted red curves, respectively, while the total mass loss rate is depicted by the dashed blue one.
Before the onset of RLO, the mass loss due to the stellar wind dominates the orbital evolution of the NS+He star system, resulting in an orbital expansion. With the onset of Case BB RLO, the mass loss rate via RLO is several orders of magnitude higher than this due to stellar winds. During the mass transfer, the mass growth of the NS is ignorable because $|\dot{M}_{\rm He}|\gg\dot{M}_{\rm Edd}$. The mass loss of the binary system plays a crucial role in influencing the orbital evolution, which can simultaneously result in two opposing effects: orbital expansion and orbital shrinkage (\citealt{tau06}).
On the one hand, the mass loss directly reduces the total gravitational force and leads to an orbital expansion;
on the other hand, the AM loss due to the mass loss causes an orbital shrinkage. The orbital evolution fates of the system would depend on the competition between these two effects.
Before the onset of RLO, the stellar winds escapes from the surface of the He star, carrying away its specific AM. Because of a relatively low mass loss rate and small AM loss rate, the first effect dominates the orbital evolution, resulting in orbital expansion.
After the onset of RLO, most (about $99\%$) of the transferred material via the Roche Lobe is ejected from the vicinity of the NS with large specific AM.\footnote{ If the NS can accrete most of the transferred material from the He star, the conservative mass transfer model predicts a negative period derivative when the mass is transferred from the more massive donor star to the less massive NS.} As a consequence, the second effect dominates the orbital evolution, causing a significant orbital shrinkage, till ${\rm lg}(t_\ast-t)\sim3.0$ in Sys. A [ ${\rm lg}(t_\ast-t)\sim2.5$, and ${\rm lg}(t_\ast-t)\sim2.0$ for Sys. B and C, respectively]. Similar to He star mass, in the following evolution, the orbital period has little change, although the absolute derivative of it is still large.

Evolution of various core masses and the total mass of the He star in Sys. A is plotted in Fig. 3.
Similar to Fig. 2, the gray dot-dashed, dashed, and dotted vertical lines correspond to the onset of Case BB RLO,
the start points of stage 2, and 3, respectively.
As shown by olive solid and wine dotted curves, the discrepancy between the evolutionary tracks of the C core and the O core is tiny,
especially in the late evolutionary stages.
For simplicity, the C core and the O core are commonly treated as a single entity, that is the CO core, in the literature.
Furthermore, it develops the neutron-rich (n-rich) core (wine solid line) at the final stage of O burning, preceding the formation of the Fe core (dashed blue), which result from a neutronization induced by the EC process in small-core stars (\citealt{arn74,thi85}).

Fig. 4 demonstrates the evolution of the central temperature ($T_{\rm c}$) of the He star with the central density ($\rho_{\rm c}$).
The case BB RLO (marked as wine triangle) initiates at ${\rm lg}(T_{\rm c}/{\rm K})\sim8.8$ and ${\rm lg}(\rho_{\rm c}/{\rm g\,cm^{-3}})\sim 5.5$, at which C is fused into Ne and/or Mg. Stage 2 starts (wine circle) at ${\rm lg}(T_{\rm c}/{\rm K})\leq9.0$, prior to the ignition that Ne is fused into Mg,
and ceases at ${\rm lg}(T_{\rm c}/{\rm K})\cong9.85$ and $\rho_{\rm c}\sim 10^{10}{\rm ~g~cm}^{-3}$. At ${\rm lg}(T_{\rm c}/{\rm K})\sim9.3$ and ${\rm lg}(\rho_{\rm c}/{\rm g\,cm^{-3}})\sim 7.0$, the He star develops a Si core (the violet pentagon).
In the stage 2, the He stars complete the majority of the nucleosynthesis, and the core mass approach the maximum values (see also Fig. 3). In the stage 3 (start from the wine square), the rapid shrinkage of the central core causes the central temperature and central density to increase by about $1-2$ magnitude within an evolution time of only 3-4 minutes because all fuels are exhausted.

\begin{figure}[h]
   \centering
   \includegraphics[trim=10 250 100 30, width=\textwidth, angle=0]{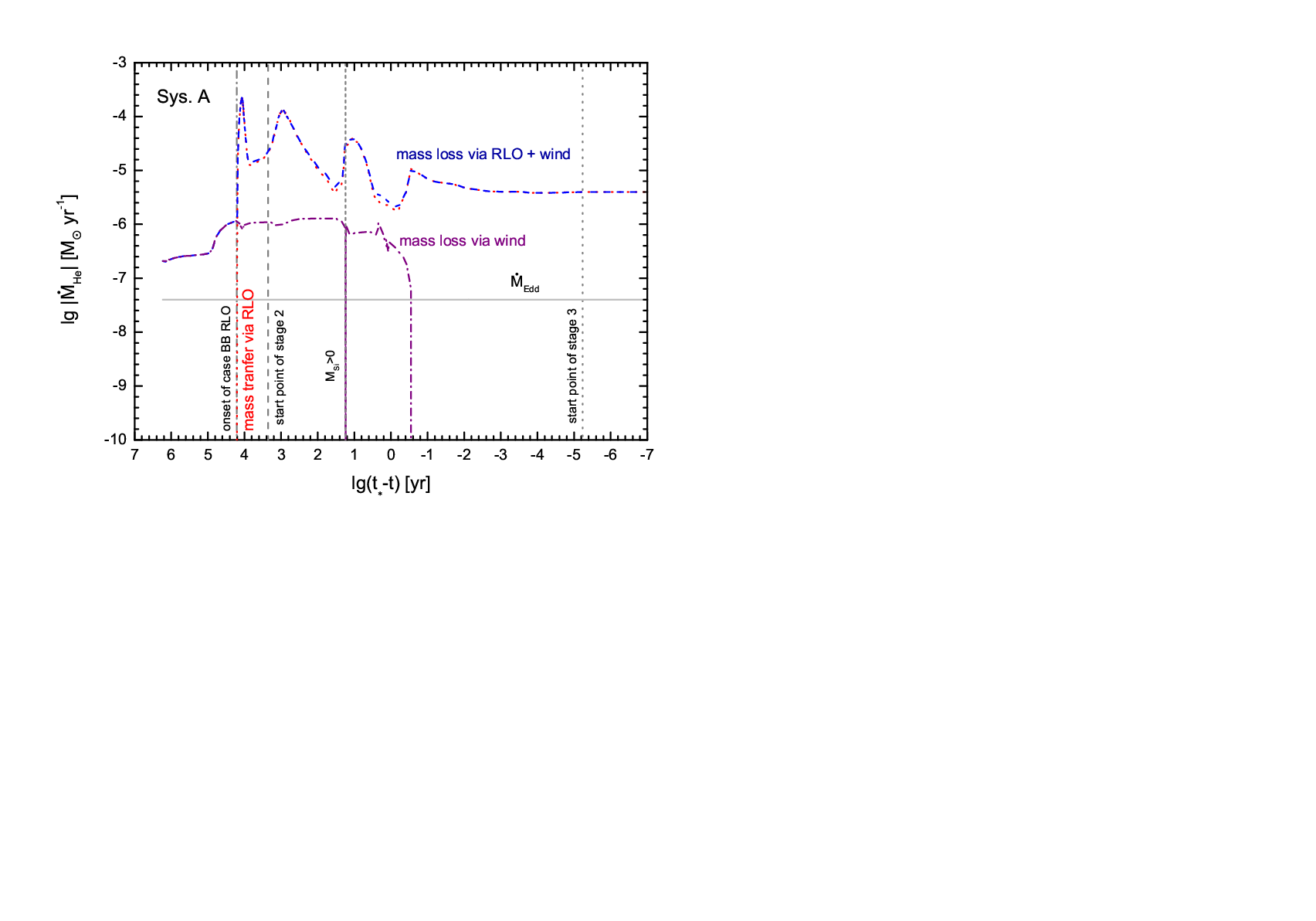}
   \caption{Evolution of mass loss rate of the He star in Sys. A. The red dotted, purple dot-dashed, and blue dashed curves illustrate the mass loss rate due to RLO, stellar wind, and RLO$+$wind, respectively. The gray dot-dashed, dashed, and dotted vertical lines correspond to the onset of Case BB RLO, the start points of stage 2, and 3, respectively. The horizontal solid gray line represents the Eddington accretion rate of the NS ($\dot{M}_{\rm Edd}=4.0\times10^{-8}~M_\odot\,\rm yr^{-1}$), while the vertical gray short dashed lines marked with '$M_{\rm Si}>0$' indicates the formation of Si core.}
   \label{Fig2}
   \end{figure}

\begin{figure}[h]
   \centering
   \includegraphics[trim=10 250 100 30, width=\textwidth, angle=0]{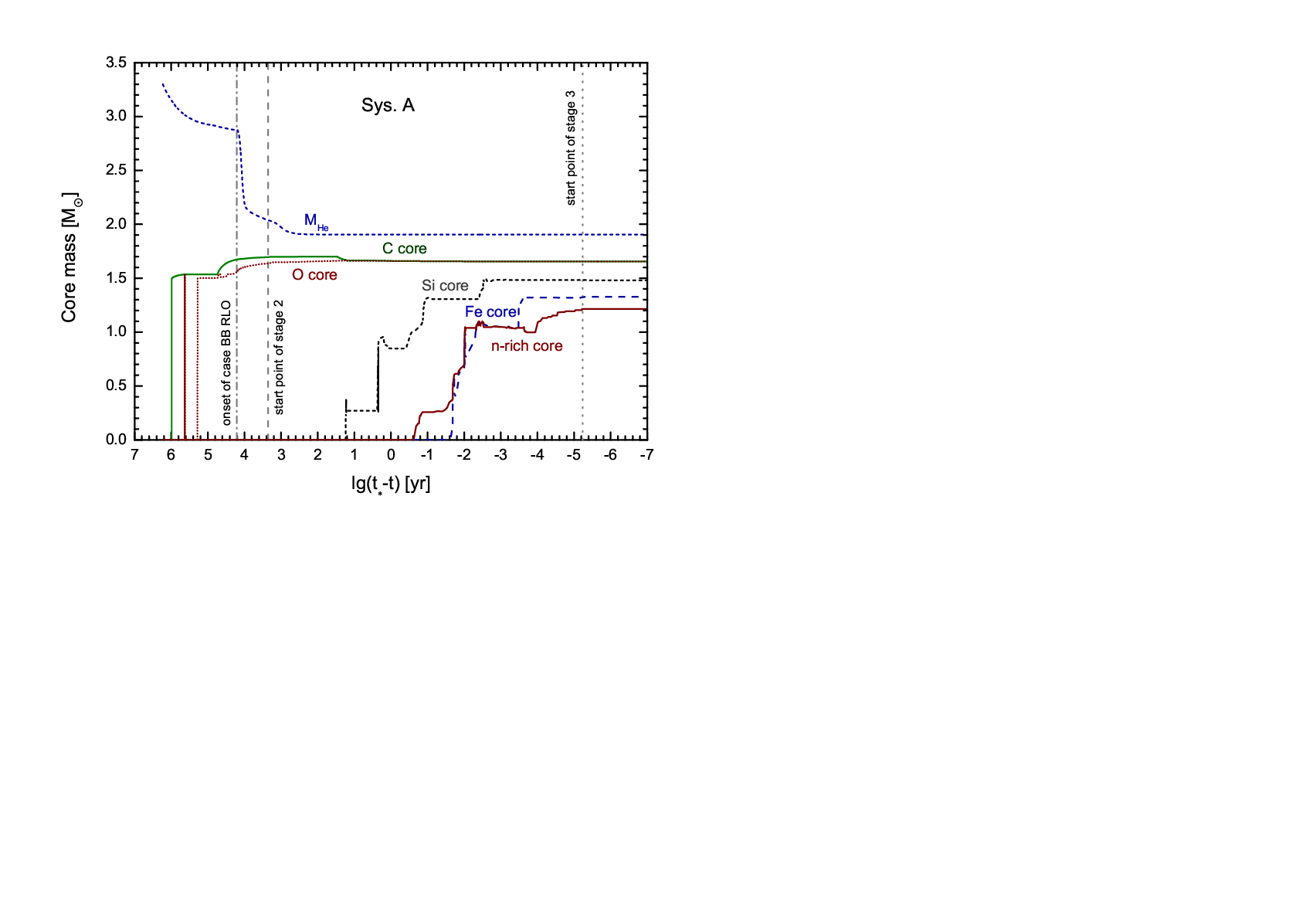}
   \caption{Evolution of various core masses (and the total mass, $M_{\rm He}$, short dashed blue line) of the He star in Sys. A. The C, O, Si, Fe, and neutron rich  cores are demonstrated by solid olive, short dotted wine, short dashed black, dashed blue, and solid wine lines, respectively. The gray dot-dashed, dashed, and dotted vertical lines correspond to the onset of Case BB RLO, the start points of stage 2, and 3, respectively. }
   \label{Fig3}
   \end{figure}

\begin{table}[h]
\begin{center}
\caption[]{Final Core masses of the Donor He Stars and SN Parameters}\label{Tab:publ-works}
 \begin{tabular}{clllllllllll}
  \hline\noalign{\smallskip}
Sys    & $M_{\rm He,f}$   &  $M_{\rm C}$  &  $M_{\rm O}$  &  $M_{\rm Si}$  &  $M_{\rm Fe}$  &  $M_{\rm {n-rich}}$ &  $M^{\rm SN}_{\rm {Ni56}}$  &  $M^{\rm bary}_{\rm NS,2}$ &   $M^{\rm grav}_{\rm NS,2}$ &  $E_{\rm SN}$\\
        & ${\rm M}_\odot$  & ${\rm M}_\odot$   & ${\rm M}_\odot$  & ${\rm M}_\odot$  & ${\rm M}_\odot$  & ${\rm M}_\odot$  & ${\rm M}_\odot$ & ${\rm M}_\odot$   & ${\rm M}_\odot$   & $10^{51}\rm~erg$ \\
  \hline\noalign{\smallskip}
A   &  1.90 &	1.66 & 1.66 &1.48 & 1.33	& 1.21     & 0.02   &1.49  &1.34 &0.58       \\ 
B   &  2.08 &	1.74 & 1.74 &1.45 & 1.44	& 1.23     & 0.03   &1.49  &1.34 &0.70     \\
C   &  2.63 &	2.10 & 2.03 &1.58 & 1.36    & 1.23     & 0.05   &1.55  &1.39 &0.93     \\
  \noalign{\smallskip}\hline
\end{tabular}
\end{center}
\end{table}

\begin{figure}[h]
   \centering
   \includegraphics[trim=10 250 100 30, width=\textwidth, angle=0]{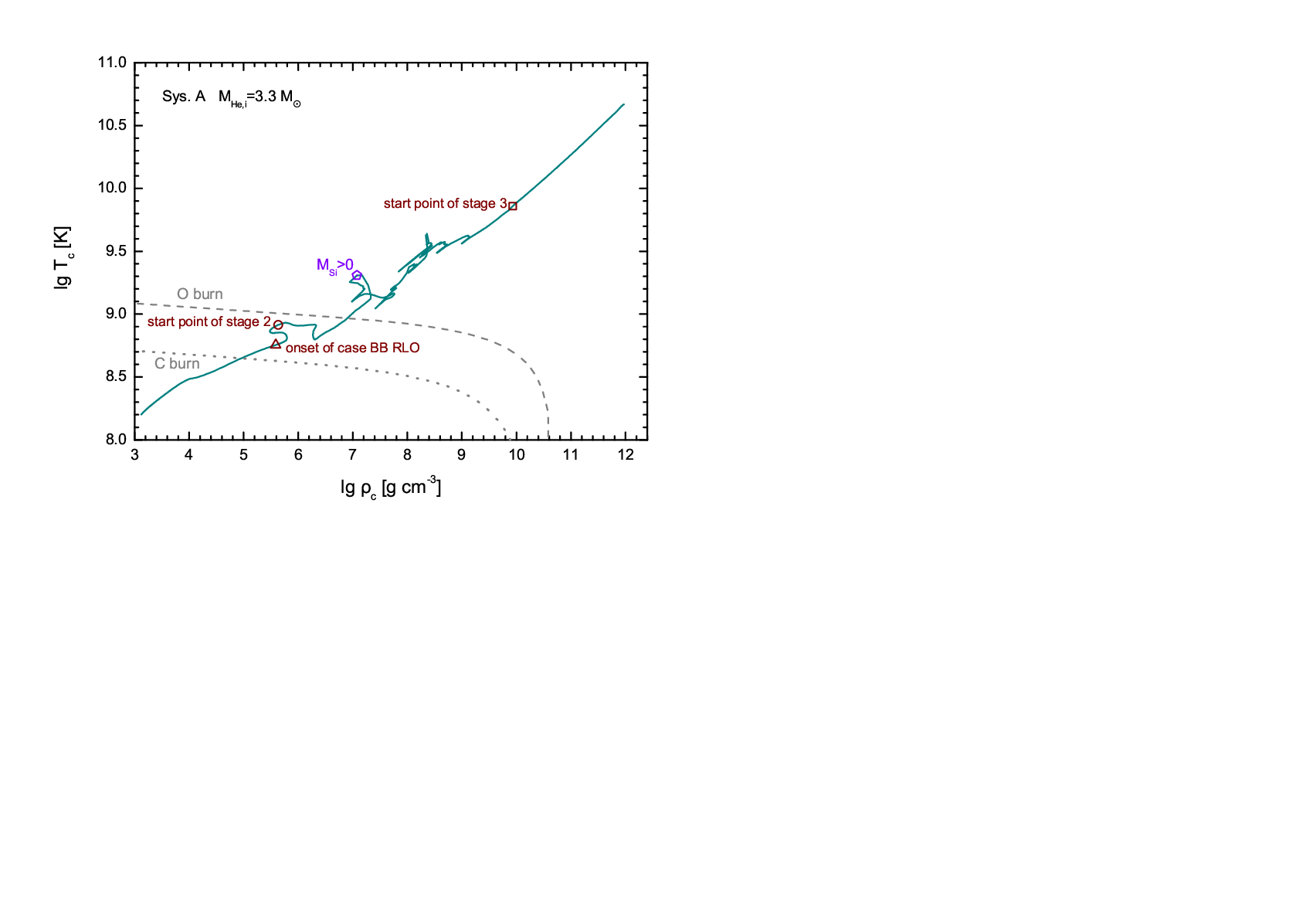}
   \caption{Evolution of central temperature with central density of the He stars in Sys. A (Dark cyan solid line). The wine triangle, circle, square, and violet pentagon denotes the onset of Case BB RLO, the starting points of stages 2, and 3, and the formation of Si core, respectively. The gray doted line and dashed line illustrated the ignition conditions of C and O, respectively.}
   \label{Fig4}
   \end{figure}

\begin{figure}[h]
   \centering
   \includegraphics[trim=10 250 100 30, width=\textwidth, angle=0]{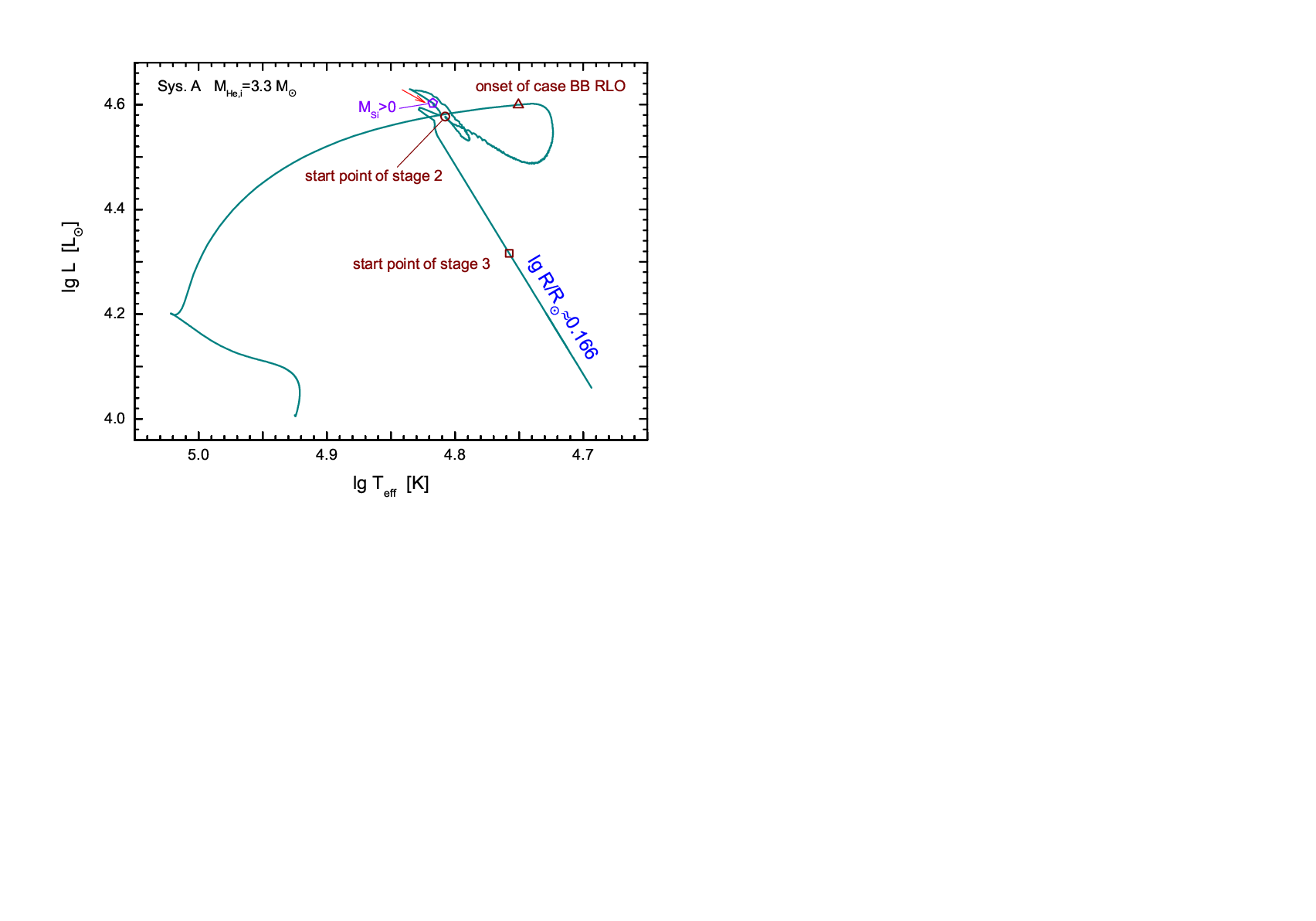}
   \caption{Evolution in the H-R diagram of the He stars in Sys. A (Dark cyan solid line). The wine triangle, circle, square denote the onset of Case BB RLO, the starting points of stages 2, and 3, respectively. The red arrow indicates the evolution direction before the formation of Si core which is marked as violet pentagon.}
   \label{Fig5}
   \end{figure}

The evolution of the He star for Sys. A in Hertzsprung-Russell diagram is displayed in Fig. 5.
At stellar age of $t\sim1.63{\rm~Myr}$ (${\rm~lg}(t_*-t)\sim4.9$), the central helium of the He star is exhausted and the surface temperature reaches a maximum ($\sim10^{5.02}$ K, corresponding to a luminosity of $\sim10^{4.2}{\rm~L}_{\odot}$). With the helium shell burning, the He star climbs the giant branch and expands,
which causes an increase of the luminosity and a decrease of the effective temperature till the onset of RLO (at $t\sim1.69{\rm~Myr}$, and ${\rm~lg}(t_*-t)\sim4.2$).
However, the central densities and temperatures of the He star steadily rise, as shown in Fig. 4.
During the RLO stage, the thermal equilibrium is disrupted due to an extremely high mass loss rate.
To keep a hydrostatic equilibrium and compensate for the mass loss of the envelope, and continue the nucleosynthesis in the inner,
the He star experiences a complex evolutionary tracks in the HR diagram and central density-temperature map until the end of Stage 2.
Such a complex evolution results in the fluctuation of the mass loss rate in Fig. 2. Here we discuss the fluctuation related to at ${\rm lg}(t_\ast-t)\sim1.2$ as an instance.

In Fig. 2, the vertical gray short dashed line marked with '$M_{\rm Si}>0$' at ${\rm lg}(t_\ast-t)\sim1.2$ illustrates the emerge of Si core (see also Fig. 3). There exist an obvious fluctuation of the mass loss rates before and after the Si core forms. In Fig. 4, there is a steady increase of central density and temperature before the formation of Si core,
which indicates the shrinkage of inner core. According to Virial theorem, half of the gravitational energy is released to heat the outer part, resulting in an expansion of the stellar envelope and a decrease of the effective temperature, which are illustrated by the slope and evolution track direction before the violet pentagon in Fig. 5. The expansion of the stellar envelope naturally causes the increase of the mass loss rate via RLO, while the decline of the effective temperature results in a decrease of the mass loss via stellar winds.

As denoted in Fig. 5, the He star evolves along the straight line with a constant radius of ${\rm lg}(R/{\rm R_{\odot}})\approx0.166$ in the final stage of the evolution. However, the central iron core experiences a rapid contraction (see also Fig. 4), which would give rise to an efficient release of the gravitational energy. A constant He-star radius implies that the released gravitational energy is lost
in the form of neutrinos, and is difficult to deposit in the outer layers.

\section{Discussion}

\subsection{Post-SN Parameters}
Based on the profile of the He stars at the moment of iron core collapse, we use the semianalytic SN explosion model of \cite{mul16} to estimate the mass of the newborn NS and other characteristics of the explosion.
As listed in Table 3, the newborn NSs in Sys. A and B are predict a same baryon mass $M^{\rm bary}_{\rm NS,2}=1.49~{\rm M}_\odot$,
but Sys. C possesses a massive NS with a baryon mass of $1.55{\rm M}_\odot$.
According to their core masses and the element profiles given in Fig. 6,
these three baryon masses imply the baryon mass cut of the newborn NS to be close to the inner edge of the Ne-rich layer.
For Sys. A and C, the baryon mass cut corresponds to the outer edge of the Si core,
that is the boundary of the Si-dominated layer.

The gravitational mass $M^{\rm grav}_{\rm NS,2}$ listed in Table 3 is calculated
according to the relation (\citealt{lat89}):
\begin{equation}
M^{\rm bary}_{\rm NS,2}\backsimeq M^{\rm grav}_{\rm NS,2}+0.084{\rm M}_\odot(M^{\rm grav}_{\rm NS,2}/{\rm M}_\odot)^2.
\end{equation}
Our predicted gravitational mass of the newborn NS is slightly higher
than the lower limit ($1.285{\rm M}_\odot$) of the companion of J1864,
implying our detailed stellar models may reproduce the evolution of its progenitor.
To fit the observed total mass of $2.63{\rm M}_\odot$, the initial NS masses in our simulation should be finely-tuned to be $\sim1.29{\rm M}_\odot$ in Sys. A and B, $\sim1.24{\rm M}_\odot$ in Sys. C.
However, these tiny difference in the initial NS masses is negligible in influencing the aforementioned binary evolution and
the subsequent dynamical processes of SN kicks.
Table 4 summarized three binary parameters prior to the SN explosions, which can be used to simulate the dynamical processes of SN kicks.

\begin{table}[h]
\begin{center}
\caption[]{Pre-SN Parameters Used to Estimate the Dynamical Effect of SN}\label{Tab:publ-works}
 \begin{tabular}{clllllllllll}
  \hline\noalign{\smallskip}
Sys & $M_{\rm NS,1}$  & $M_{\rm He,f}$   & $P_{\rm orb,f}$  & $M_{\rm NS,2}$   \\
    & ${\rm M}_\odot$ & ${\rm M}_\odot$  &  day             & ${\rm M}_\odot$   \\
  \hline\noalign{\smallskip}
A  & 1.35  & 1.9  & 0.41  & 1.34	    \\ 
B  & 1.35  & 2.1  & 0.42  & 1.34        \\
C  & 1.35  & 2.6  & 0.40  & 1.39	    \\
  \noalign{\smallskip}\hline
\end{tabular}
\end{center}
\end{table}

According to the study of \cite{hil83},
the ratio between the semi-major axes before and after SN is (\citealt{hil83,dew03,sha16}):
\begin{equation}
  \frac{a_0}{a} = 2-\frac{M_{\rm NS,1}+M_{\rm He, f}}{M_{\rm NS,1}+M_{\rm NS, 2}}[1+(\frac{V_{\rm K}}{V_0})^2+2\frac{V_{\rm K}}{V_0}{\rm cos}\theta]
\end{equation}
where $V_{\rm K}$ and $V_0$ are the kick velocity and the pre-SN orbital velocity of the He star, respectively, while $\theta$ is the angle between them.
Assuming that the positional angle of $V_{\rm K}$ with respect to the pre-SN orbital plane is $\varphi$, the post-SN eccentricity satisfies (\citealt{hil83,dew03,sha16}):
\begin{equation}
  1-e^2 = \frac{a_0}{a}\frac{M_{\rm NS,1}+M_{\rm He, f}}{M_{\rm NS,1}+M_{\rm NS, 2}}[1+(\frac{V_{\rm K}}{V_0})^2({\rm cos}^2\theta+{\rm sin}^2\theta{\rm sin}^2\varphi)+2\frac{V_{\rm K}}{V_0}{\rm cos}\theta]
\end{equation}

To determine the distribution of post-SN orbital periods and eccentricities,
we model the dynamical processes due to SN explosions with two fixed kick velocities for Sys. A, B, and C.
We assume a random (isotropic) kick direction, which corresponds
to a uniform distribution of ${\rm cos}\theta=-1$ to 1 and
$\varphi= 0-\pi$.
As demonstrated in the top panel of Fig. 6, the wide distribution of three simulated systems can encompasses
the observed value of J1846 when $v_{\rm k}=100{\rm~km~s}^{-1}$. However,
only Sys. A may be the progenitor of J1846 for a relatively small kick velocity of $50{\rm~km~s}^{-1}$.

\begin{figure}[h]
   \centering
   \includegraphics[trim=10 250 100 30, width=\textwidth, angle=0]{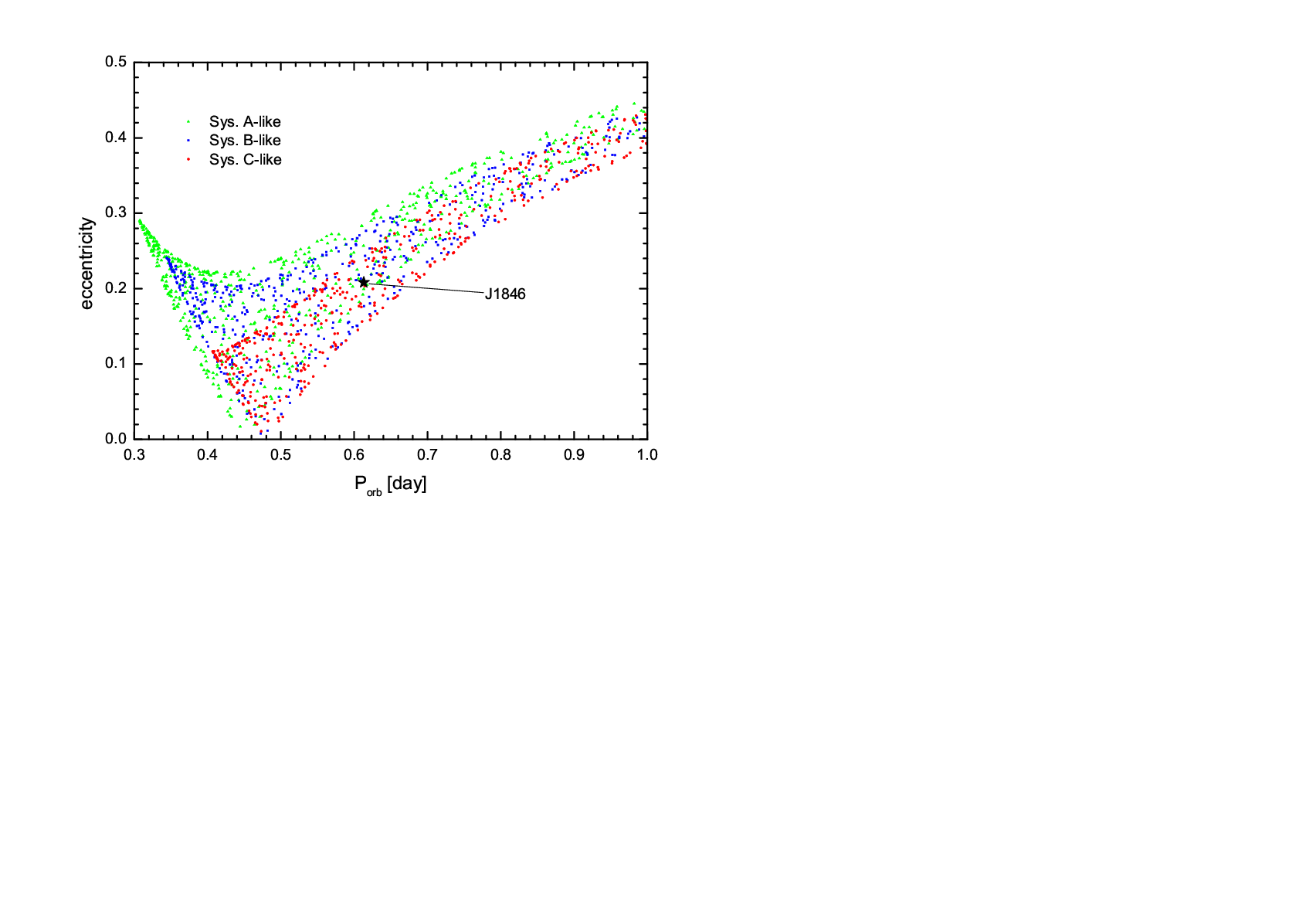}
   \includegraphics[trim=10 250 100 30, width=\textwidth, angle=0]{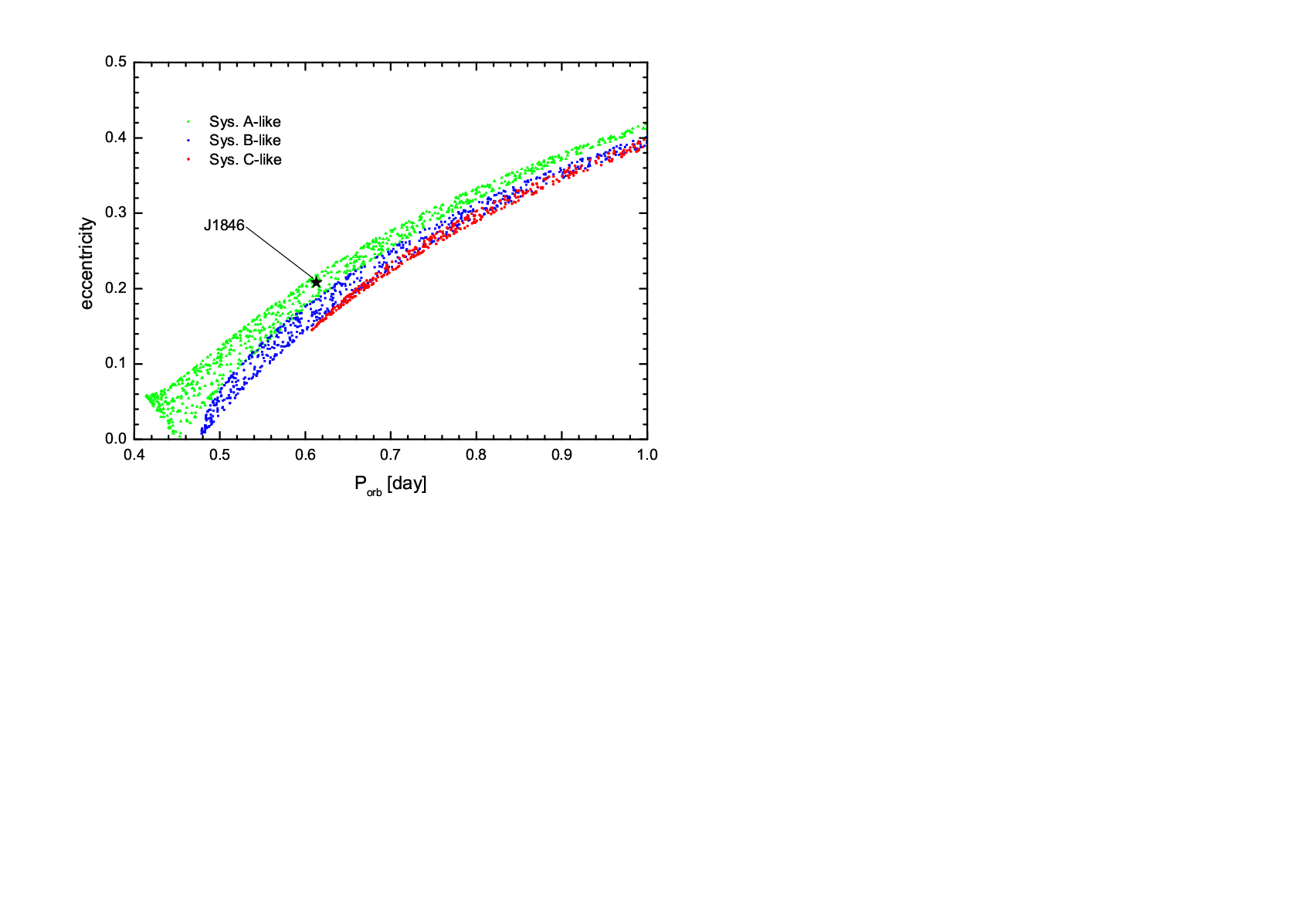}
   \caption{Distribution of post-SN DNSs in the eccentricity vs. orbital period diagram for systems A-like (green triangle), B-like (blue square), and C-like (red circle). The black solid star represents the observational parameters of J1846.
   The kick velocities in the top panel and bottom panel are $V_{\rm K}=100{\rm~km~s}^{-1}$ and $50{\rm~km~s}^{-1}$, respectively.   }
   \label{Fig6}
   \end{figure}

\subsection{Progenitors of DNS Merge GW Sources}
Close DNS systems are anticipated to merge within a Hubble time
and appear as high-frequency GW sources detected by ground-based detectors, such as LIGO, Virgo, and KAGRA.
The first high-frequency GW event originated from a DNS merger is GW170817, which provided the first multi-messenger astronomical observations that including gravitational waves, radio, infrared, optical, ultraviolet, X-ray, and gamma-ray detection (\citealt{abb17}). The merger time of eccentric DNS with an initial orbital semimajor axis $a$ and eccentricity $e_0$ can be calculated based on the loss of orbital energy due to GW radiation (\citealt{tau17}):
\begin{equation}
  \tau_{\rm GWR} = \frac{15c^5}{304G^3}\frac{a^4 C_1^4}{M_{\rm NS,1}M_{\rm NS, 2}(M_{\rm NS,1}+M_{\rm NS, 2})}\times f(e)
\end{equation}
where $C_1$ is a constant of initial eccentricity,
\begin{equation}
C_1=\frac{1-e_0^2}{e_0^{12/19}}[1+(121/304)e^2_0]^{-870/2299}
\end{equation}
while
\begin{equation}
f(e)=\int_0^{e_0}\frac{e^{29/19}[1+(121/304)e^2]^{1181/2299}}{(1-e^2)^{3/2}}{\rm d}e
\end{equation}
is an integration function of eccentricity which cannot be solved analytically (except when $e_0=0$).
Adopting the observed parameters of J1846, i.e. $M_{\rm NS,1}=1.3455$, $M_{\rm NS, 2}=1.2845$,  $P_{\rm{orb}}=0.613\rm{~day}$, and $e=0.208$, one can get a merge time of $\sim8.7$ Gyr.

Setting an upper limit of 13.8 Gyr for the merger time,
we calculate the merger possibilities of post-SN simulated systems under different kick velocities.
As shown in Fig. 7, three post-SN systems have a merger probability of $\gtrsim30\%$ for a kick velocity range of $V_{\rm K}\sim(100-400){\rm~km~s}^{-1}$. For Sys. A, the merge probability increases to $\gtrsim50\%$ when $V_{\rm kick}\lesssim150{\rm~km~s}^{-1}$. When the kick velocity approaches zero, the merger times of Sys. A, B, and C converge to constants of 9.4 Gyr, 14.0 Gyr, and 23.6 Gyr, respectively. As a result, the difference of the merger times leads to a rapid increase of the merge probability in Sys. A and a decrease of those in Sys. B and C in the low kick range.

\begin{figure}[h]
   \centering
   \includegraphics[trim=10 250 100 30, width=\textwidth, angle=0]{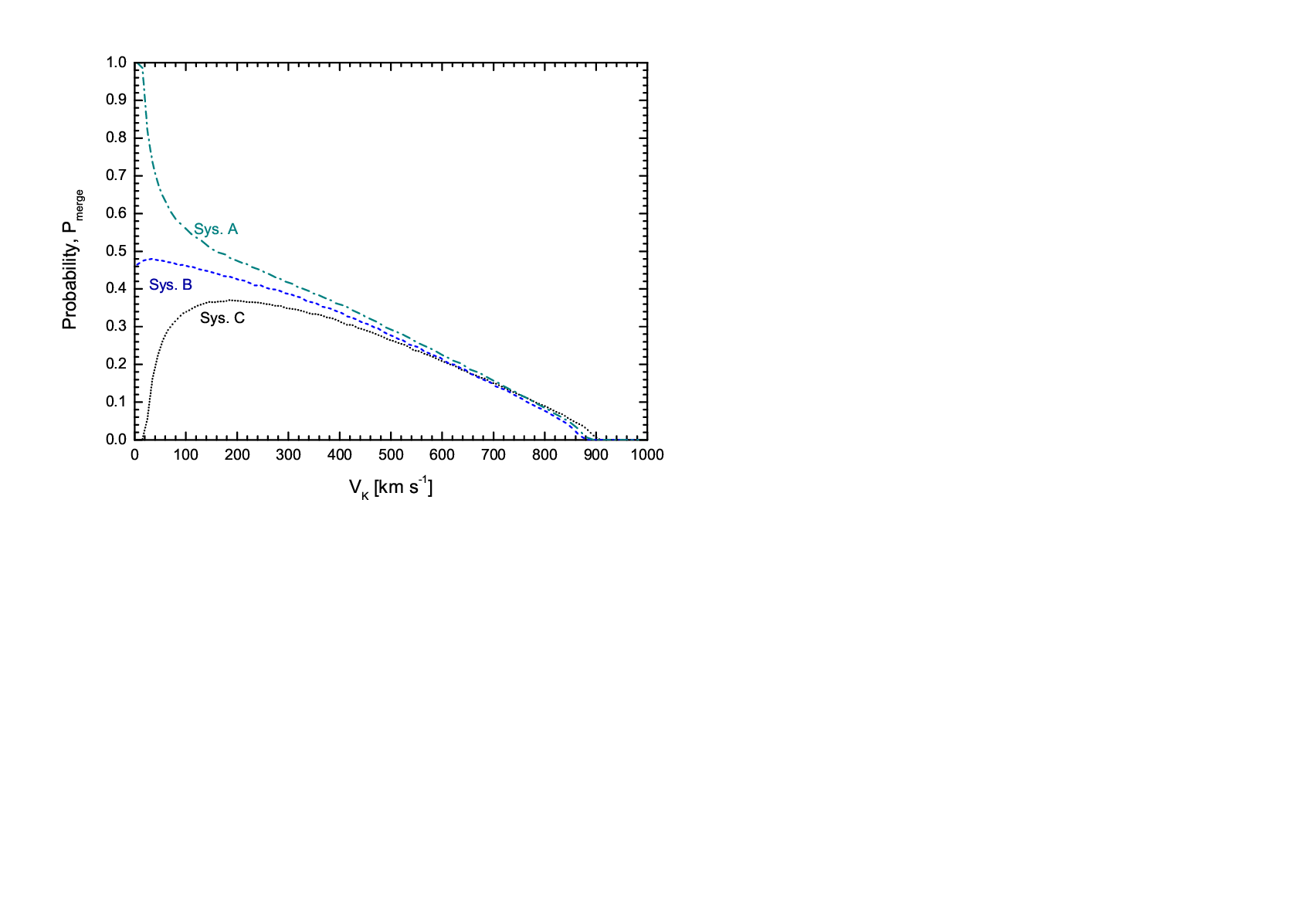}
   \caption{Merger probabilities of post-SN DNSs within a Hubble time (13.8 Gyr) for Sys. A (dot-dashed dark cyan line), B (short dashed blue line), and C (short dotted black line). }
   \label{Fig7}
   \end{figure}

\subsection{EC-SN progenitor}
Induced by the EC on $^{24}{\rm Mg}$ and $^{20}{\rm Ne}$, an oxygen$-$neon (ONe) core
with a mass close to the Chandrasekhar limit will collapse into a NS via an EC-SN explosion (\citealt{nom84,nom87,jon13,doh15,tau15,tau23}).
Guo et al. (2024) conducted a systematic study of EC-SNe in NS+He star binaries,
and constrained the ONe core masses produced EC-SNe to be a narrow range from 1.385 to $1.43{\rm ~M}_\odot$.
Based on the mass of the companion star of J1846, the ONe core mass prior to EC-SN is at least $1.424{\rm ~M}_\odot$ according to  Eq. (1). As a consequence, the ONe core mass is in a narrow range of $1.424\leq M_{\rm ONe}\leq1.43{\rm ~M}_\odot$ if the newborn NS is produced by an EC-SN. Therefore, the He star mass must be finely tuned if the companion star of J1846 stems from an EC-SNe.

\section{Conclusions}
\label{sect:conclusion}
Since the binary parameters including the mass ratio, the distributions of spin period and orbital period, and the relationship between orbital period and eccentricity are consistent with those of the recycled pulsars in known DNS systems, the recycled pulsar J1846 was proposed to be a DNS (\citealt{zhao24}). To test whether a DNS like J1846 can be formed via isolated binary evolution, we simulate the detailed binary evolution of three binaries consisting of a NS and a He star with an initial mass of 3.3, 3.5, and $4.0{\rm ~M}_\odot$, in which the initial orbital periods are 0.55, 0.50, and 0.50 days, respectively. Our calculations found that three helium stars can evolve from the He ZAMS stage to the iron core collapse prior to Fe CCSN, and eventually form newborn NSs.

Employing the semianalytic SN explosion model given by \cite{mul16},
we derive the post-SN parameters including the masses of the newborn NS and the explosive kinetic energies.
Considering the dynamical effect of SN kick, we calculate the distribution of DNS systems in the orbital period versus eccentricity ($P_{\rm orb}-e$) plane. Our models can successfully reproduce
the observational parameters of J1846 when the kick velocity is $100{\rm~km~s}^{-1}$.
For kick velocities in the range of $100<V_{\rm kick}<400{\rm~km~s}^{-1}$,
the simulated post-SN systems will merge with a probability of $\gtrsim30\%$ in a Hubble time,
producing a GW event similar to GW170817.

\begin{acknowledgements}
We are grateful to the anonymous reviewer for constructive and detailed comments that improved our
manuscript. This work was partly supported by the National Natural Science Foundation of China
(under grant Nos. 12373044, 12273014, 12203051, 12403035, 12393811, 12288102, and 12041304),
the Natural Science Foundation (under grant Nos. ZR2023MA050, and ZR2021MA013)
of Shandong Province, the China Postdoctoral Science Foundation (under grant Nos. 2024M751375 and 2024T170393),
the Postdoctoral Fellowship Program of CPSF (under grant No. GZB20240307),
the Jiangsu Funding Programme for Excellent Postdoctoral Talent (under grant No. 2024ZB705),
the Tianshan Talent Program of Xinjiang Uygur autonomous region (under grant No. 2023TSYCTD0013).
and the CAS 'Light of West China' Program (under grant No. 2018-XBQNXZ-B-022).
\end{acknowledgements}




\label{lastpage}

\end{document}